\documentclass[aps,prl,twocolumn,superscriptaddress,english]{revtex4}
\usepackage{amssymb}
\usepackage{graphicx}
\usepackage{amsmath}
\usepackage{soul}
\usepackage{amsthm}
\usepackage{amsfonts}
\usepackage[T1]{fontenc}
\usepackage[latin9]{inputenc}
\usepackage{array}
\usepackage{multirow}
\usepackage{color}
\usepackage{esint}
\usepackage{bm}
\usepackage{color}
\usepackage{bbm}
\usepackage{hyperref}
\usepackage{babel}
\usepackage{titlesec}

\begin{document}

\global\long\def\id{\mathbbm{1}}
\global\long\def\ui{\mathbbm{i}}
\global\long\def\ud{\mathrm{d}}

\title{Many-body critical phase: extended and nonthermal}

\author{Yucheng Wang}
\affiliation{Shenzhen Institute for Quantum Science and Engineering, and Department of Physics,
Southern University of Science and Technology, Shenzhen 518055, China}
\affiliation{International Center for Quantum Materials, School of Physics, Peking University, Beijing 100871, China}
\affiliation{Collaborative Innovation Center of Quantum Matter, Beijing 100871, China}
\author{Chen Cheng}
\affiliation{School of Physical Science and Technology, Lanzhou University, Lanzhou 730000, China}
\author{Xiong-Jun Liu}
\thanks{{Corresponding author: xiongjunliu@pku.edu.cn}}
\affiliation{International Center for Quantum Materials, School of Physics, Peking University, Beijing 100871, China}
\affiliation{Collaborative Innovation Center of Quantum Matter, Beijing 100871, China}
\affiliation{CAS Center for Excellence in Topological Quantum Computation, University of Chinese Academy of Sciences, Beijing 100190, China}
\affiliation{Shenzhen Institute for Quantum Science and Engineering, and Department of Physics,
Southern University of Science and Technology, Shenzhen 518055, China}
\author{Dapeng Yu}
\affiliation{Shenzhen Institute for Quantum Science and Engineering, and Department of Physics,
Southern University of Science and Technology, Shenzhen 518055, China}

\begin{abstract}
The transition between ergodic and many-body localization (MBL) phases lies at the heart in understanding quantum thermalization of many-body systems. Here we predict a many-body critical (MBC) phase with finite-size scaling analysis in the one-dimensional extended Aubry-Andr\'{e}-Harper-Hubbard model, which is different from both the ergodic phase and MBL phase, implying that the quantum system hosts three different fundamental phases in the thermodynamic limit. The level statistics in the MBC phase are well characterized by the so-called critical statistics, and the wave functions exhibit deep multifractal behavior only in the critical region. We further study the half-chain entanglement entropy and thermalization properties, and show that the former in the MBC phase manifest a volume law scaling, while the many-body states violate eigenstate thermalization hypothesis. The main results are confirmed with the state-of-art numerical calculations. This work unveils a novel many-body phase which is extended but nonthermal.
\end{abstract}
\maketitle

{\em Introduction.---} In the past decade, the eigenstate thermalization
hypothesis (ETH) \cite{Huse3,Deutsch,Rigol,Deutsch2} has become an essential theoretical underpinning
for understanding quantum thermalization physics. Eigenstates in an ergodic phase obey the ETH, while including disorder in interacting systems
can lead to the many-body localization (MBL) if the disorder strength is strong enough. The MBL phase
violates ETH, i.e., the states in MBL cannot thermalize. The existence of MBL phases has been
well established in one-dimensional interacting systems with random disorder \cite{Basko,Huse1,Huse2,Huse3,Altman,Huse4,Varma,Ioffe2019} or incommensurate potential \cite{Iyer,Modak,Li,Li2,Wang,Lee,Chandran,Nag,Gray,Setiawan,Crowley,Hong,Mirlin2019}, and has also been observed in interacting ultracold atomic gases trapped in incommensurate optical lattices \cite{Bloch1,Bordia,Bloch2,Bloch3,Bloch4}.

The nature of the transition from an ergodic phase to MBL remains an active area of research. The entanglement entropy (EE) of eigenstates in an ergodic phase follows the volume law, but in MBL obeys an area law \cite{Pollmann,Grover,Bauer}.
In transition between such two phases, the EE changes in a singular way, rendering an eigenstate phase transition. On the other hand, in view of the long-time dynamics in the ergodic phase and MBL phase \cite{Vosk,Serbyn,Serbyn2,Prosen,Bardarson,Lea1},
a dynamical phase transition is manifested between the two phases.
The critical features of the transition are also examined in the recent works~\cite{Setiawan,Pollmann,Grover,Agarwal,Pekker,Vasseur,Lea2,HuseX,Ioffe,Rispoli} to reveal the nature of the critical points.
In particular, an outstanding question is, whether there exists some sort of critical phase, other than the critical point in the phase transition? The results based on the finite-size analyses showed that a quantum critical region in the finite-size interacting system vanishes in the thermodynamic limit~\cite{Setiawan,HuseX,Luitz,GG,Kudo,Potter,Potter2,Dumitrescu,Laflorencie}, for which the emergence of a many-body critical (MBC) phase is yet illusive. To confirm the existence of such critical phase which is different from both the ergodic and MBL phases is undoubtedly
important in understanding the quantum thermalization physics of many body disorder systems.

In this work, we show that a MBC phase can exist in thermodynamic limit in an extended Harper model \cite{Hatsugai,Han,Takada,Chong} with Hubbard interaction. Employing exact diagonalization (ED) to obtain various diagnostics such as level statistics, multifractality, EE and thermalization properties, we show with finite-size scaling analysis the existence of this new phase of
quantum matter which is different from the ergodic and MBL phases, and is confirmed with the state-of-art numerical study. The level statistics is shown to follow the critical statistics \cite{Shore,Nicopoulos,Moshe,Muttalib,Nishigaki,GG2} and the wave functions exhibit deep multifractal behaviors. Moreover, the eigenstates in this critical phase violate the ETH but their EE follow a volume law, so this MBC phase is an extended nonthermal phase.

\begin{figure}
\centering
\includegraphics[width=0.5\textwidth]{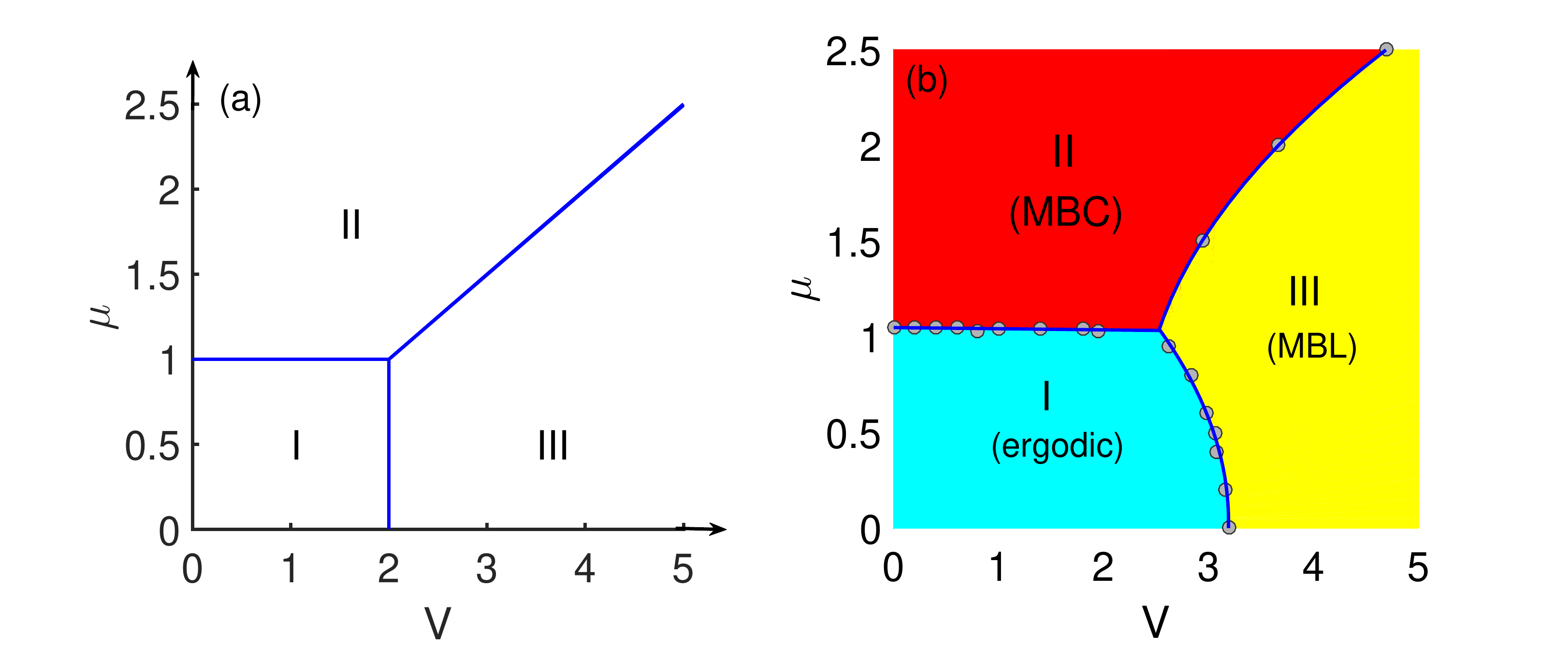}
\caption{\label{01} Main results.
(a) Non-interacting phase diagram. The regions I, II, and III correspond to extended, critical, and localized phases, respectively. (b) Phase diagram for the extended interacting Harper model with $U=1$, which contains three phases: the ergodic phase (region I), the MBL phase (region III) and the MBC phase (region II). The transition points (marked by gray dots) can be obtained with finite-size scaling analyses for EE and level statistics, and the phase boundaries are fitted by corresponding curves. Only in the MBC phase of region II, the level statistics follow the critical statistics and the many-body wave functions are multifractal.}
\end{figure}

{\em Model and phase diagram.---}
We start with the extended Harper model from the Hamiltonian~\cite{Hatsugai,Han,Takada,Chong}
\begin{eqnarray}\label{ham-1}
H_0 &=& \sum_{j}\{(1+\mu \cos[2\pi(j+\frac{1}{2})\alpha+\delta])c^{\dagger}_{j}c_{j+1}+H.c.\nonumber\\
 &+& V\cos(2\pi j\alpha+\delta)c^{\dagger}_{j}c_{j}\}.
\end{eqnarray}
where $c_{j}$ ($c^{\dagger}_j$) is the fermion annihilation (creation) operator at the site j, $\mu$ represents the amplitude of the
modulation in the off-diagonal hopping, V is the strength of the
on-site incommensurate potential and $\delta$ is an arbitrary phase shift. We take $\alpha=\frac{\sqrt{5}-1}{2}$, then both the on-site potential and the hopping amplitude between the nearest-neighboring (NN) lattice sites are incommensurate.
Fig. \ref{01}(a) shows the phase diagram of this system \cite{Hatsugai,Han,Takada,Chong}, where the region I, II and III correspond to the single-particle extended, critical, and localized phases respectively.
When $\mu=0$, this model is reduced to the Aubry-Andr$\acute{e}$-Harper model \cite{AA} and $V=2$ is the transition point from the extend to the localized phases.

We then add the NN repulsive Hubbard interaction and the total Hamiltonian is described by
\begin{equation}\label{ham}
H=H_0+U\sum_{j} n_{j}n_{j+1},
\end{equation}
where $n_j = c^{\dagger}_jc_j$ is the fermion number operator and $U$ represents the interaction strength. We consider the half-filling case, with the numbers of fermions $N$ and the lattice sites $L$ being fixed to $N/L = 1/2$.
Since the sample-to-sample fluctuations in quasiperiodic models are much weaker than those in random disorder models, the number of samples used ranges from $500$ to $10$ for our study, where a sample is specified by choosing an initial phase $\delta$. We obtain the eigenstates by ED under open boundary condition, and, unless otherwise stated, focus on the states in the middle one-third of the spectrum for $L\leq18$ and middle $100$ states for bigger sizes (see Supplementary Material (SM)~\cite{Supplemental}).
The main results are shown in Fig.~\ref{01} (b), where the interaction is taken $U=1$ as an example. Three fundamental phases, i.e., the ergodic phase (I), the MBL (III), and the MBC phase (II) which constitutes the most important prediction in this work, are uncovered. We show below the phase diagram and further explore the fundamental properties of the MBC phase.

{\em Energy level statistics and multifractal analysis.---}
The MBC phase in the region II in Fig.~\ref{01}(b) can be identified from the energy level statistics and multifractal behavior. Defining the energy spacing as $\delta_n=E_{n+1}-E_{n}$, where the eigenvalues $E_n$ have been listed in ascending order. Then we can obtain the ratio of adjacent gaps as $r_n$=$\frac{\min(\delta_n, \delta_{n+1})}{\max(\delta_n, \delta_{n+1})}$ and average it over all gaps and samples. For the system in the ergodic phase, its level statistics follow Gaussian-orthogonal ensemble (GOE): $P_G=\frac{\pi}{2}\frac{\delta}{\langle\delta\rangle}\exp(-\frac{\pi\delta^2}{4\langle\delta\rangle^2})$, where $\langle\delta\rangle$ is the mean spacing and $\langle r\rangle$ converges to $0.529$. In the MBL phase, the level statistics are Poisson: $P_P=\frac{1}{\langle\delta\rangle}\exp(-\frac{\delta}{\langle\delta\rangle})$ and $\langle r\rangle\approx 0.387$ \cite{Huse1,Huse2}.
One can see $P_P$ is maximum at $\delta=0$ while in ergodic phase $P_G(\delta=0)=0$. The latter corresponds to the level repulsion of spectra in the ergodic phase.
The larger $\langle r\rangle$ in the ergodic phase tells that the spectrum of the ergodic phase is more uniform than that in the MBL phase.
However, the value of $\langle r\rangle$ in the region II of Fig. \ref{01} (b) is neither $0.387$ nor $0.529$ (see more details in SM~\cite{Supplemental}), implying that the level statistics are neither GOE nor Poisson.

\begin{figure}
\centering
\includegraphics[height=65mm,width=92mm]{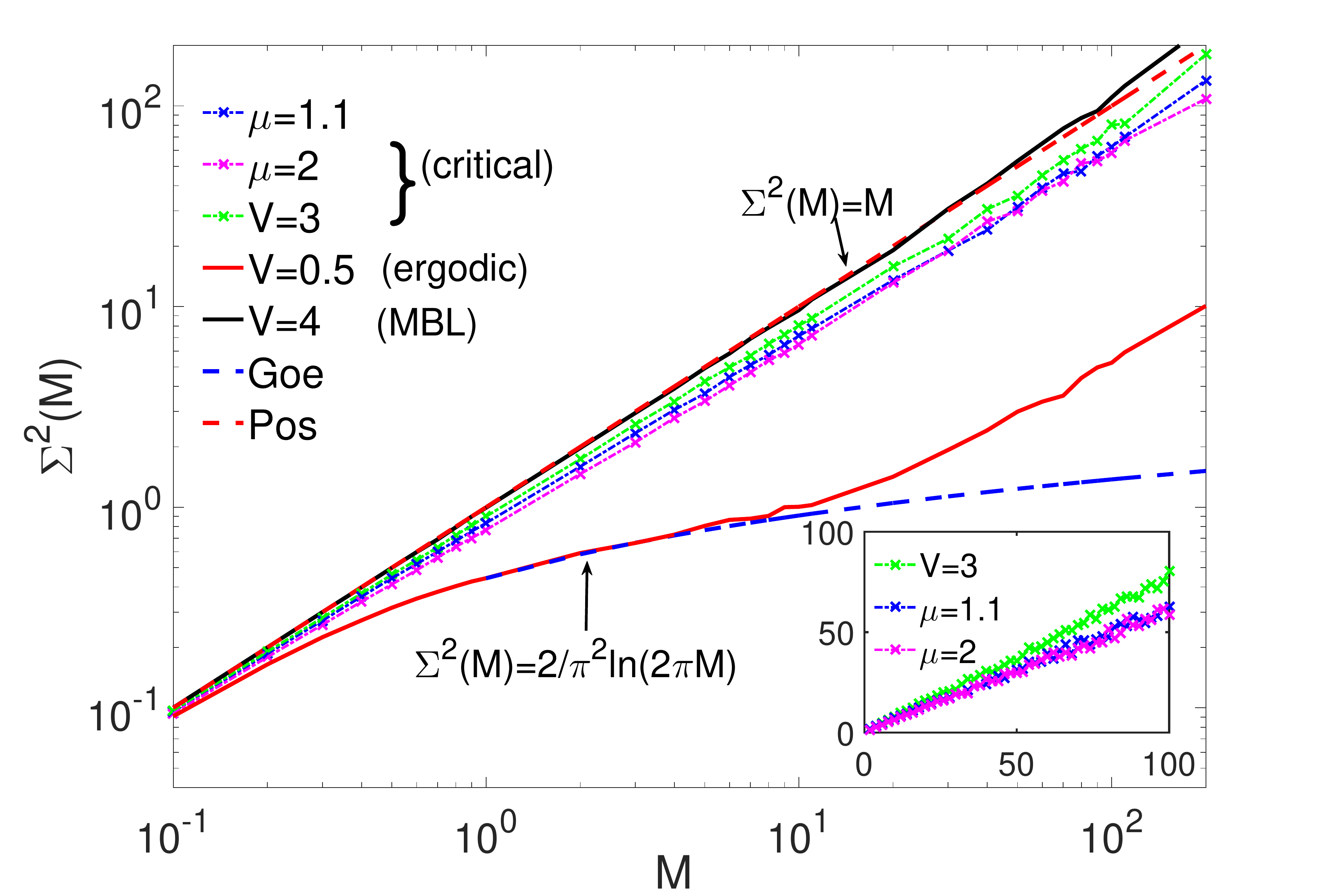}
\caption{\label{04}
Number variance $\Sigma^2(M)$ for different incommensurate potential strengths V with fixed $\mu=0.5$, and for different $\mu$ with fixed $V=1$. In the ergodic phase and when $M$ is small, $\Sigma^2(M)$ shows a slow logarithmic increase (red solid curve). In the MBL phase, $\Sigma^2(M)$ is linear with slope one (black solid curve). Inset: Number variance in the critical regime and near the transition point. They are linear with slopes $1/2 < \chi < 1$, which is a signature of critical statistics. Here we fix the number of sites $L=16$.}
\end{figure}

To characterize the statistical properties of the energy spectra in the region II, we consider the level number variance $\Sigma^2(M)$, which is given by:
$\Sigma^2(M(\epsilon))=\langle M^2(\epsilon)\rangle-\langle M(\epsilon)\rangle^2$,
with $\langle M(\epsilon)\rangle$ counting the number of levels in a strip of width $\epsilon$ on the unfolded scale~\cite{Dyson,Guhr,GG}. The unfolding procedure is using a smooth function to fit the staircase function $\eta(E)=\sum_{m}\Theta(E-E_m)$, which counts the number of eigenvalues less than and equal to $E$ (see SM \cite{Supplemental}). The angular bracket denotes the average over different regions of the mid one-third spectrum and different samples.
In the MBL phase, the spectrum has no correlations, and therefore the number variance is exactly linear with slope one, i.e, $\Sigma^2(M)=M$ (black solid curve in Fig.~\ref{04}).
The spectrum in the ergodic phase, as mentioned above, is more uniform due to the level repulsion, so the number variance displays a slow logarithmic growth: $\Sigma^2(M)\approx\frac{2}{\pi^2}\ln(2\pi M)$ (red solid curve in Fig. \ref{04}). When $M$ is bigger, the number variance in the ergodic phase shows a much faster power-law growth, which is considered to be a evidence of the existence of the Thouless energy \cite{GG,Thouless1}.
The number variance in the MBC phase 
is qualitatively different, which is linear $\Sigma^2(M)\sim \chi M$ but with a slope less than one $\chi < 1$, as given in Fig. \ref{04}. In order to see it clearly, we redisplay the number variance of the critical regime in the inset and confirm that the number variances are asymptotically
linear with slopes $1/2 < \chi < 1$. The slopes are size-insensitive and signify the critical statistics~\cite{Supplemental}. The critical statistics can be used to describe the energy spectrum of a high-dimensional, non-interacting disordered system near the extended-Anderson localization transition point. The number variance in the MBC phase is intermediate between the ergodic phase and the MBL phase, which means that the uniformity and the strength of level repulsions of the spectrum are also in between.


\begin{figure}
\hspace*{-0.3cm}
\includegraphics[width=0.5\textwidth]{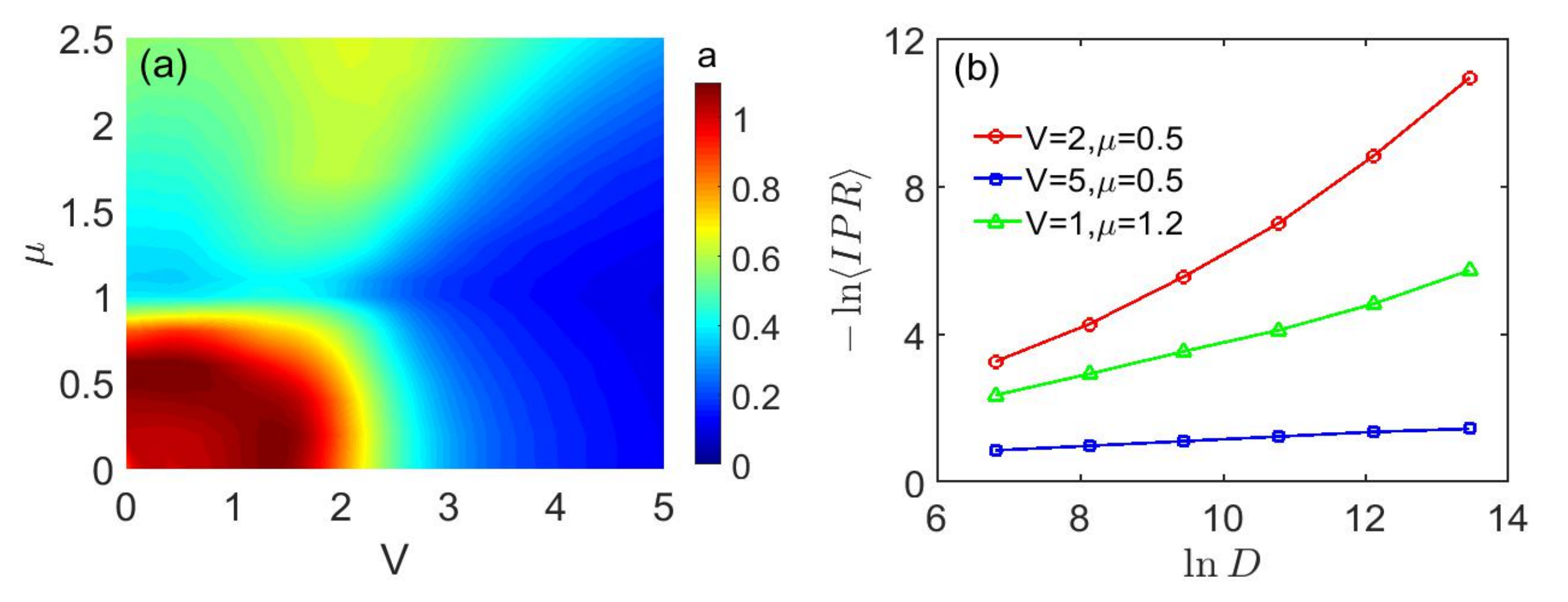}
\caption{\label{05}
 (a) Fractal dimension $a$ as a function of $V$ and $\mu$. (b) -$\ln \langle IPR\rangle$ as a function of $\ln D$ for different system sizes from $L=12$ to $L=22$. The fitting coefficients are $a=1.11\pm 0.15, b=-3.48\pm 1.02$ for $V=2, \mu=0.5$ (in the ergodic phase), $a=0.0514\pm 0.0668, b=0.21\pm 0.04$ for $\mu=0.5, V=5$ (in the MBL phase, the scope of $a$ includes $0$) and $a=0.51\pm 0.08, b=-0.77\pm 0.28$ for $V=1, \mu=1.2$ (in the MBC phase). Here we consider the mid $100$ states for all sizes.}
\end{figure}

We further show the MBC phase through the multifractal analysis based on the system size from $12$ to $L=22$, and the analysis can be performed by examining
the fractal dimension $a$~\cite{Bell,Luitz2014}, defined for each eigenstate
$|n\rangle$ as $a=-\ln \langle\sum_j|\psi_j|^4\rangle/\ln D$ with $D\rightarrow\infty$, where $\psi_j$ is the wave function coefficient of the eigenstate $|n\rangle$ in
the computational basis $\{|j\rangle\}$, given by $\psi_j=\langle j|n\rangle$, $\langle\sum_j|\psi_j|^4\rangle$ is the averaged inverse participation ratio (IPR) \cite{Thouless1974}, and $D={{L}\choose{N}}$ is the Hilbert space size.
Fig.~\ref{05} (a) show the fractal dimensions $a$, which is obtained by using ${\ln \langle IPR\rangle}=a\ln D+c$ to fit {$\ln \langle IPR\rangle$} and $\ln D$ with different sizes, where $c$ is a constant.
We see that $a\approx 1$ in the ergodic phase, and $a$ is near 0 (i.e. weak multifractality~\cite{Laflorencie}), which is a consequence of
the finite-size effect (see more details in SM~\cite{Supplemental}), or $a=0$ within error bars (i.e., absence of multifractal behavior) \cite{Luitz,Luitz2016,Gao2020,explain} in the MBL phase.
In contrast, the MBC phase is deeply multifractal, with the fractal dimension $a$ being finite and not close to $0$, nor $1$, manifesting a new phase different from ergodic and MBL phases. Further, more careful analysis shows a logarithmic
subleading correction to the scaling of $\langle IPR\rangle$, i.e., $-\ln\langle IPR\rangle=a\ln D+b\ln(\ln D)+o(\ln D)$. We find that $b$ is negative, negative, and positive for the ergodic, MBC, and MBL phase respectively [Fig. \ref{05} (b)], as summarized in the Table \ref{coefficients}.

\begin{table}\renewcommand{\arraystretch}{1.2}
  \centering
\begin{tabular}{|l|l|ll}
\cline{1-2}
    coefficients  &   phases & \\ \cline{1-2}
    $a$ $\approx$ 1, b < 0  &  ergodic phase & \\ \cline{1-2}
    $a$ is far from 0 and 1, b < 0 & MBC phase & \\ \cline{1-2}
    $a$ is near 0 or $a$=0, b > 0  &  MBL phase & \\ \cline{1-2}
\end{tabular}

\caption{A comparison of the coefficients $a$ and $b$ in different phases. }\label{coefficients}
\end{table}

The critical level statistics and deep multifractal behavior of many-body wave functions show that the region $II$ in Fig. \ref{01} (b) is a MBC phase, qualitatively different from both the ergodic and MBL phases. An intuitive understanding for the existence of the MBC phase with interactions is given in the SM by connecting the many-body wavefunction and single-particle orbits~\cite{Supplemental}. Next we proceed to study the EE and thermalization with finite-size scaling analyses, which confirm the critical points of phase transition in the thermodynamic limit and give characteristic features of the MBC phase.

\begin{figure}
\centering
\includegraphics[width=0.5\textwidth]{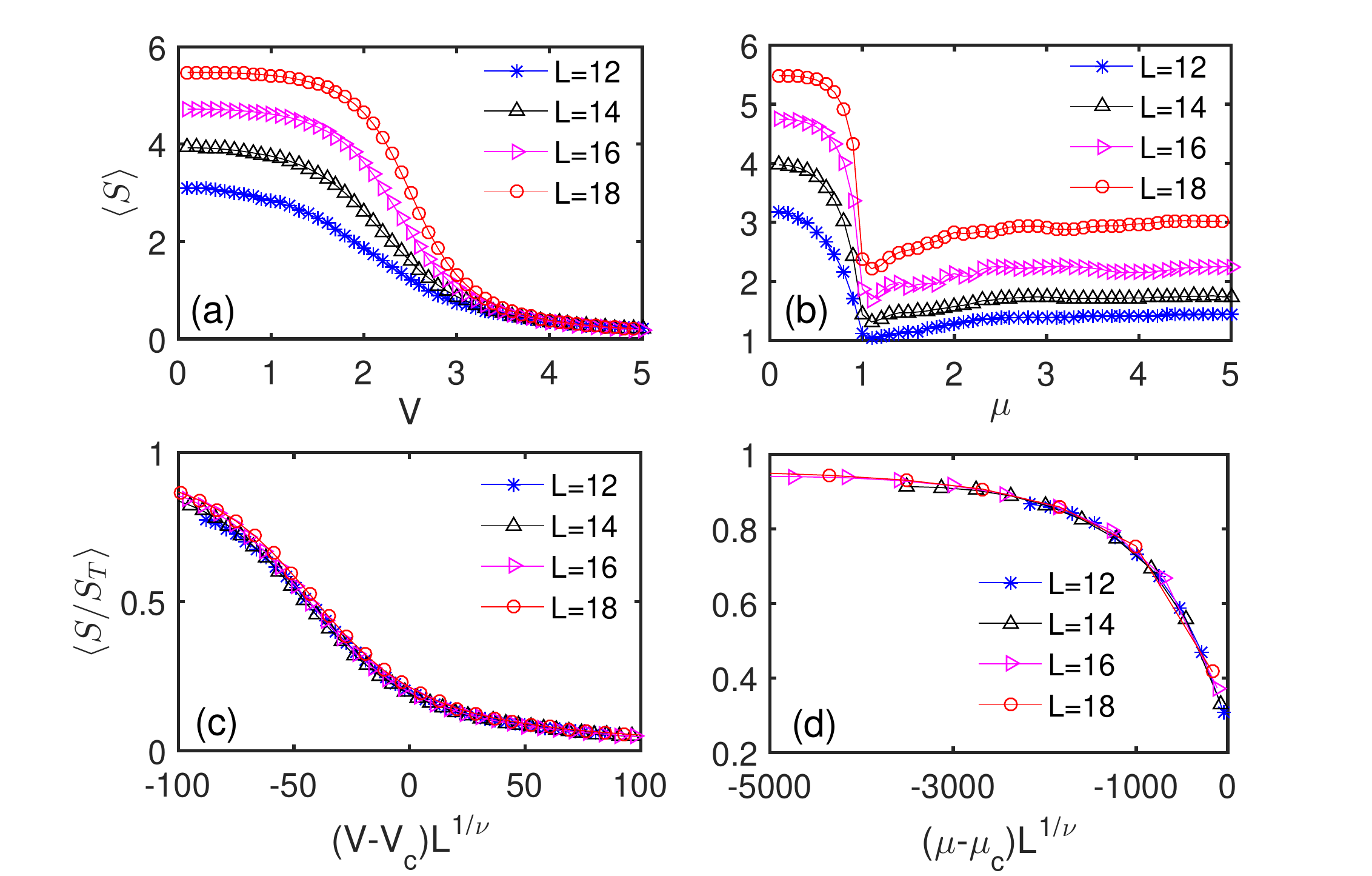}
\caption{\label{02}
 (a) The sample averaged EE $\langle S\rangle$ as a function of V with fixed $\mu=0.5$ and (b) $\langle S\rangle$ as a function of $\mu$ with fixed $V=1$. Finite-size scaling analysis of $\langle S/S_T\rangle$ as a function of $(V-V_c)L^{1/\nu}$ with fixed $\mu=0.5$ in (c) and $(\mu-\mu_c)L^{1/\nu}$ with fixed $V=1$ in (d). Rescaled $\langle S/S_T\rangle$ so that all curves for different sizes converge to a single curve. }
\end{figure}
{\em Finite-size scaling and entanglement entropy.---}The EE is an important resource to explore the critical behaviors and the localization features of many-body states based on finite-size scaling analysis. 
For a correlated system the EE is obtained by $S=-\sum_i\lambda_i\ln \lambda_i$, where $\lambda_i$ is the $i$th eigenvalue of a reduced density matrix, which can be obtained by tracing out half of this system.
In Fig. \ref{02}(a) and (b), we display the average EE $\langle S\rangle$ averaged over the mid one-third states and over samples as a function of V and $\mu$, with fixed $\mu=0.5$ and $V=1$, respectively. From Fig. \ref{02}(a), one can see that the average EE follows a volume law in the ergodic phase while decreases to a constant independent of $L$
in the deep MBL phase, which fulfills an area law. Unlike the MBL, we show in Fig.~\ref{02}(b) that the average EE follows a volume law when the system is the critical phase. For a more precise study, we perform a finite-size scaling analysis for the EE which is rescaled by the Page value $S_T=0.5[L\ln2-1]$~\cite{HuseL,Page}, as shown in Fig.~\ref{02}(c) and (d), where the results are fit to $\langle S/S_T\rangle=f[(V-V_c)L^{1/\nu}]$ with fixed $\mu=0.5$, and $\langle S/S_T\rangle=g[(\mu-\mu_c)L^{1/\nu}]$ with fixed $V=1$, respectively. Here $V_c$ and $\mu_c$ denote the transition point from the ergodic phase to MBL phase and MBC phase, respectively, and $\nu$ is the associated critical exponent. In confirming the existence of phase transition it is sufficient to perform the finite-size analysis only in one side of the critical points, say with $\mu<\mu_c$. The critical point is further obtained with the state-of-art numerical study on the largest system size $L=22$ by the shift invert diagonalization method (see details in the SM~\cite{Supplemental}).
The results of the best fit are (c) $V_c = 3.09 \pm 0.05$ and $\nu = 0.71 \pm 0.06$ and (d) $\mu_c = 1.03 \pm 0.02$ and $\nu = 0.32 \pm 0.03$, which are double confirmed with the same results obtained by performing the finite-size scaling analyses for energy level statistics~\cite{Supplemental}.
The critical exponent $\nu$ can be determined similarly for generic parameters $V$ and $\mu$ near the phase boundaries, 
and exhibits a variation range in the whole phase boundaries with different $V_c$ and $\mu_c$, which is typical for disordered systems. The approximate ranges are
\begin{eqnarray}\label{ap1}
\nu=\left\{ \begin{array}{ll}
        0.6(0)\sim 0.8(2), \ {\rm for}\ I\rightarrow III, \\
             0.3(5)\sim 0.4(7), \ {\rm for}\ I\rightarrow II, \\
             0.6(5)\sim 0.8(4), \ {\rm for}\ II\rightarrow III,
        \end{array} \right.
\end{eqnarray}
where I, II and III are the ergodic, MBC and MBL phases respectively. The scaling exponents for the transitions from ergodic to MBL phases and from critical to MBL phases are close. This is consistent with the result that the MBC phase is extended, with EE satisfying volume law and being more similar to the ergodic phase.

Note that a finite-size system may exhibit a critical region near the transition point between the ergodic and MBL phases~\cite{Agarwal,Pekker,Vasseur,Lea2,HuseX}, when the system size is smaller than the correlation length (for ergodic phase) or localization length (for MBL phase). The critical region vanishes in the thermodynamic limit after performing finite-size analyses~\cite{HuseX}. In contrast, from the finite-size scaling analysis, we confirm that the MBC phase predicted here exists in the thermodynamic limit.

\begin{figure}
\hspace*{-0.5cm}
\centering
\includegraphics[width=0.55\textwidth]{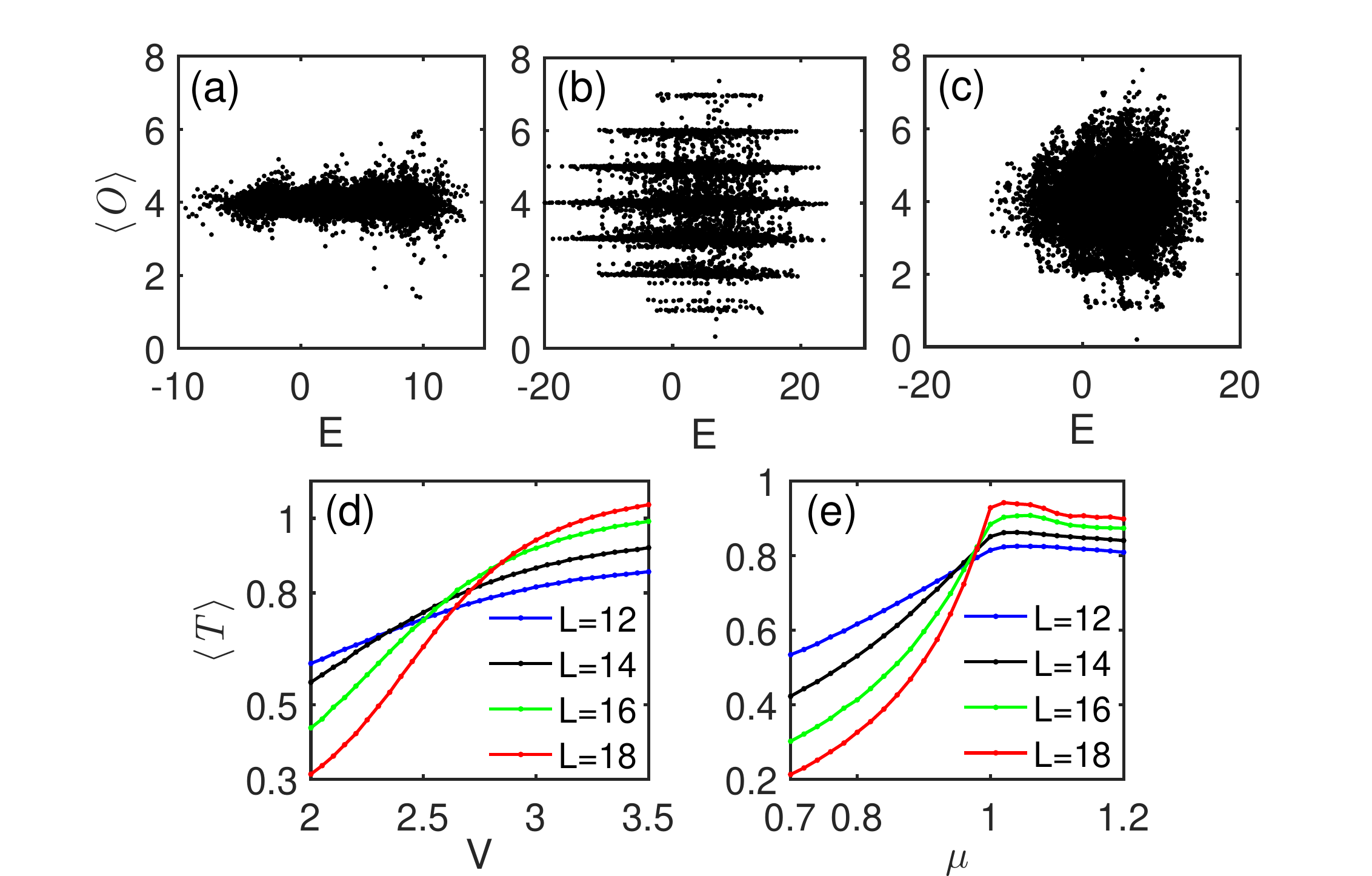}
\caption{\label{06}
 The observable $O(E)$ versus eigenvalues for (a) $V=1$, $\mu=0.5$ in the ergodic phase, (b) $V=4$, $\mu=0.5$ in the MBL phase and (c) $V=1$, $\mu=1.2$ in the MBC phase. Here the system size has been fixed as $L=2N=16$ and the initial phase $\delta=0$.  $\langle\it{T}\rangle$ as a function of (d) $V$ with fixed $\mu=0.5$ and (e) $\mu$ with fixed $V=1$ for different sizes.}
\end{figure}

{\em Thermalization properties.---} We finally study the thermalization properties of the MBC phase. For this we consider the
average deviation of the half-chain particle-number distribution from the half-number ($N/2$) of particles, which can characterize the thermalization of the system and is defined by $\it{T}=[\frac{1}{D_2-D_1}\sum_{m=D_1}^{D_2}(O(E_m)-N/2)^2]^{1/2}$, with many-body eigenstates of eigenvalue $E_m$ being summed over. Here the observable $O(E)=\sum_{j=1}^{L/2}\langle\psi_{E}|n_j|\psi_{E}\rangle$ quantifies the number of particles distributed in the half chain of the lattice for the many-body eigenstate $|\psi_{E}\rangle$ with energy $E$~\cite{Li}. The large fluctuation
of $O(E)$ among nearby eigenstates signifies the violation of the ETH.
In the ergodic phase, the fluctuations of $O$ are small and the ETH is satisfied [Fig.~\ref{06}(a)], while in the MBL phase, the fluctuations are obviously larger and the ETH is violated [Fig.~\ref{06}(b)]. For parameters in the critical regime, as shown in Fig.~\ref{06} (c), the fluctuations of $O$ are also large, which implies that the ETH is violated and the eigenstates are non-thermal. This feature can be even clearer by examining the qualitative behaviors of $T$ defined for the mid one-third of states with $D_1=\frac{D}{3}$ and $D_2=\frac{2D}{3}$.
As the system size $L$ increases, the value $O(E)$ tends to $N/2$ in the ergodic phase but keeps fluctuating and results in finite values of $T$ at large $L$ for non-thermal phase. The numerical results of the sample averaged $\langle\it{T}\rangle$ are presented in Fig.~\ref{06} (d) and (e), as a function of $V$ and $\mu$. With the increasing of system size $L$, we see that $\langle\it{T}\rangle$ decreases if the phase is ergodic, but enlarges in the MBL and MBC phases. Therefore, the critical phase is non-thermal, similar to the MBL phase. Together with the preceding discussion on the EE, we conclude that the MBC phase is an extended non-thermal phase.

{\em Conclusions.---} We have predicted that a MBC phase exists in the thermodynamic limit in the 1D extended Aubry-Andr\'{e}-Harper-Hubbard model. 
Being a third type of fundamental phase, distinct from ergodic and MBL phases, the MBC phase shows novel basic features. 
First, by analysing number variance, we found that the level statistics in the critical phase are neither GOE for ergodic regime nor Poisson for MBL, but are well described by critical statistics. Further, from a multifractal analysis 
we showed that the many-body states in the MBC phase exhibit the deep multifractal behavior. 
Finally, we unveiled that the predicted critical phase violates the
ETH but their EE exhibits a volume law, implying that this exotic critical phase is delocalized but non-ergodic and not thermal. The main results are confirmed for system of size $L=22$ by the state-of-art numerical method.
As a new interacting phase beyond the ergodic and MBL phases, many interesting issues, including the dynamical properties, deserve further efforts of study. Our work opens a door to explore quantum thermalization physics in the MBC phases.

We thank Xiaopeng Li for valuable discussions. This work was supported by National Nature Science Foundation of China (Grants No.11761161003, No.11825401, and No.11921005), the National Key R\&D Program of China (2016YFA0301604), Guangdong Innovative and Entrepreneurial Research Team Program (No.2016ZT06D348), the Science, Technology and Innovation Commission of Shenzhen Municipality (KYTDPT20181011104202253), the Open Project of Shenzhen Institute of Quantum Science and Engineering (Grant No. SIQSE202003), and the Strategic Priority Research Program of Chinese Academy of Science (Grant No. XDB28000000). Chen Cheng is supported by the National Natural Science Foundation of China (No. 11904145). The computations were partially performed in the Tianhe-2JK at the Beijing Computational Science Research Center.


\setcounter{equation}{0} \setcounter{figure}{0}
\setcounter{table}{0} 
\renewcommand{\theparagraph}{\bf}
\renewcommand{\thefigure}{S\arabic{figure}}
\renewcommand{\theequation}{S\arabic{equation}}

\onecolumngrid
\flushbottom
\newpage

\section*{\normalsize SUPPLEMENTAL MATERIAL}
In the Supplementary Materials, we first show the average ratio $\langle r\rangle$ and perform the finite size scaling for energy level statistics. Then, we provide the details in unfolding procedure and scaling analysis.
Finally, we characterize different phases from a one-particle perspective.

\section{I. Energy level statistics}

\begin{figure}[b]
\includegraphics[width=0.6\textwidth]{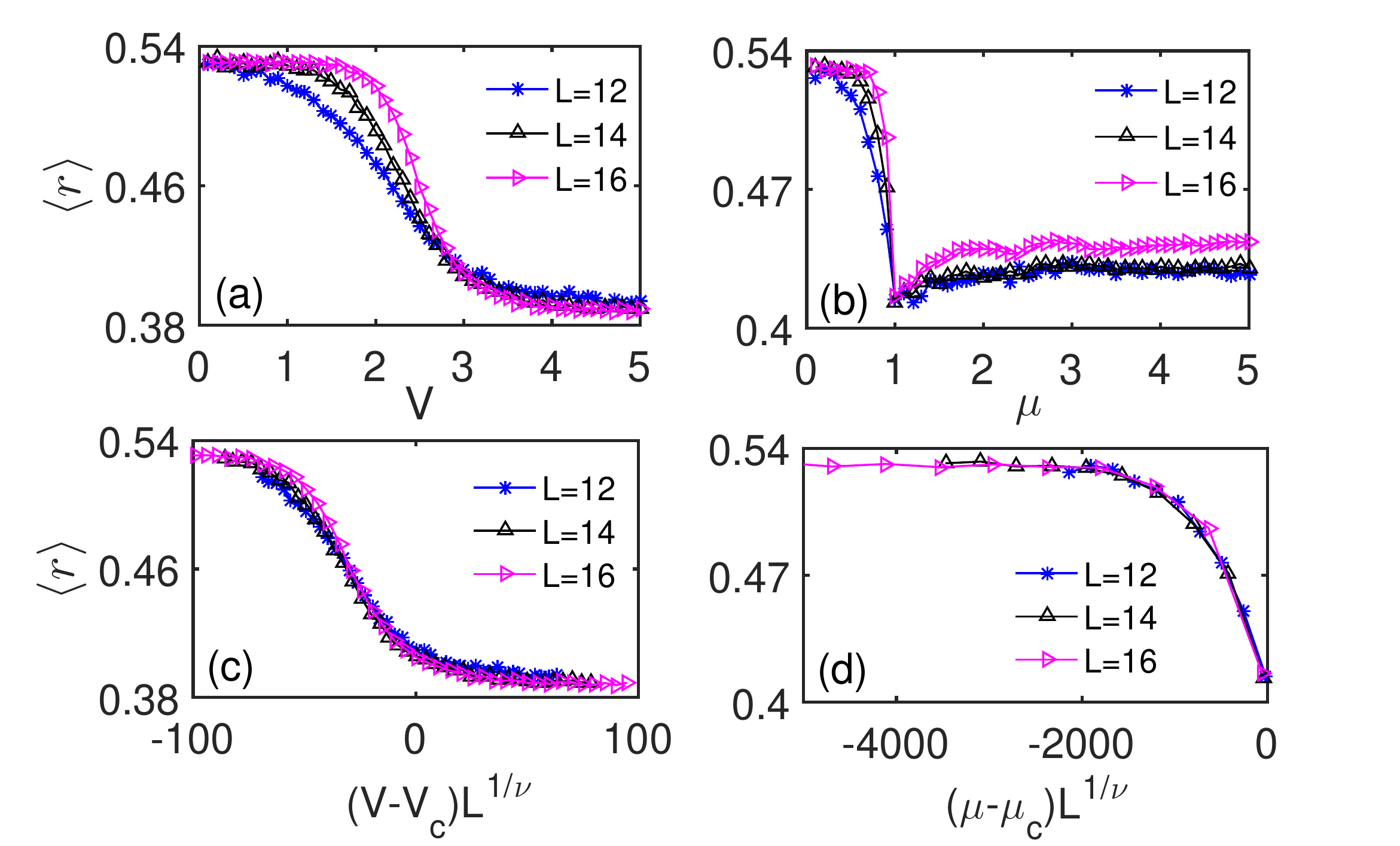}
\caption{\label{S1}
The average ratio $\langle r\rangle$ of adjacent energy gaps versus (a) $V$ with fixed $\mu=0.5$ and (b) $\mu$ with fixed $V=1$. Finite-size scaling analysis of $\langle r\rangle$ versus $f[(V-V_c)L^{1/\nu}]$ with fixed $\mu=0.5$ in (c) and $g[(\mu-\mu_c)L^{1/\nu}]$ with fixed $V=1$ in (d). Rescaled $\langle r\rangle$ so that all curves for different sizes converge to a single curve.}
\end{figure}

When we research the energy level statistics in the main text, we define the ratio of adjacent gaps as $r_n$=$\frac{\min(\delta_n, \delta_{n+1})}{\max(\delta_n, \delta_{n+1})}$, where $\delta_n=E_{n+1}-E_{n}$, where the eigenvalues $E_n$ have been listed in ascending order.
Fig. \ref{S1} (a) displays the ratio $\langle r\rangle$ averaged over mid one-third gaps and samples as a function of $V$ with fixed $\mu=0.5$ and we see that $\langle r\rangle$ changes from $0.529$ to $0.387$ when $V$ is increased from the ergodic phase to the many body localization (MBL) phase. Fig. \ref{S1} (b) shows $\langle r\rangle$ as a function $\mu$ with $V=1$ and one can see that $\langle r\rangle$ in the critical phase is neither equal to $0.387$ nor equal to $0.529$, which indicates that the level statistics are neither Gaussian-orthogonal ensemble (GOE) nor Poisson. As discussed in the main text, the value of
$\langle r\rangle$ presents the uniform degree of a spectrum. We see that $\langle r\rangle$ in the critical phase is intermediate between the ergodic and MBL phases, signifying that the uniformity of the spectrum are also between the two, which is agreement with the conclusion obtained by using the level number variance. A general energy level spacings distribution is derived as $P(\delta_n)=C_1\delta^{\beta}_nexp(-C_2\delta_n^{2-\gamma})$~\cite{Serbyn2016}, where $C_{1,2}$ are constants, $0\leq\beta\leq 1$ and $\gamma$ respectively control the level repulsion and the tail of distribution. The Poisson and GOE distribution respectively correspond to $\gamma=1, \beta=0$ and $\gamma=0, \beta=1$. Different parameters in the many body critical (MBC) phase should give different
$\gamma$ and $\beta$, and so $\langle r\rangle$ in the MBC region should distribute on a scale without tending to a unique value.
We further perform the finite-size scaling for the average ratio $\langle r \rangle$ \cite{Kudo2018}.
Fig. \ref{S1} (c) and (d) show $\langle r\rangle$ fit to a form $f[(V-V_c)L^{1/\nu}]$ with fixed $\mu=0.5$ and $g[(\mu-\mu_c)L^{1/\nu}]$ with fixed $V=1$ respectively, where $V_c$ and $\mu_c$ denote the transition point from the ergodic phase to MBL phase and the MBC phase respectively, and $\nu$ is the associated critical exponent. The results of the best fit are (c) $V_c = 3.09 \pm 0.07$ and $\nu = 0.72 \pm 0.09$ and (d) $\mu_c = 1.01 \pm 0.03$ and $\nu = 0.32 \pm 0.03$, which is consistent with the results obtained by performing the finite-size scaling analysis for the entanglement entropy in the main text.

\begin{figure}[t]
\centering
\includegraphics[width=0.5\textwidth]{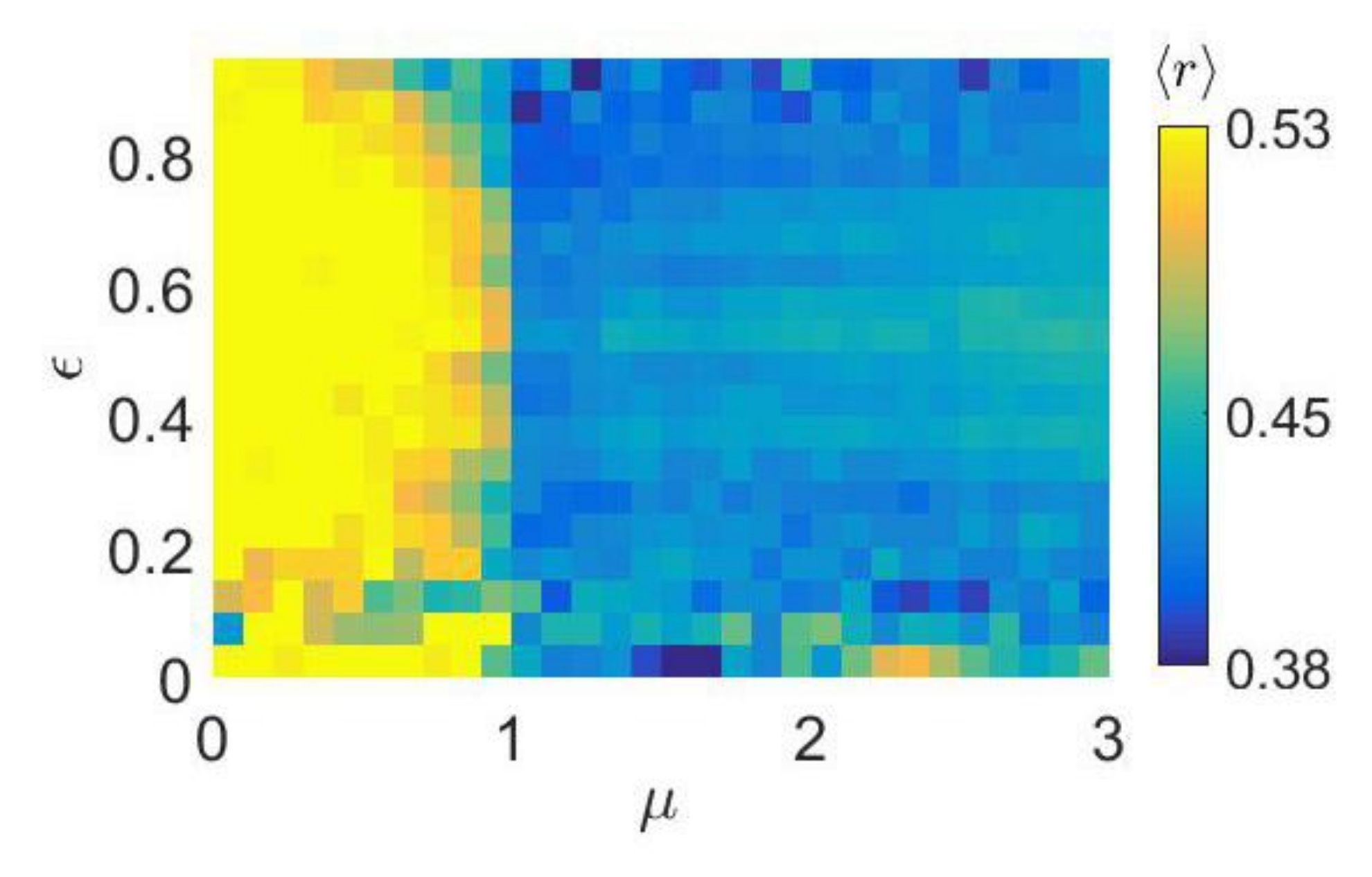}
\caption{\label{me}
The averaged $\langle r\rangle$ versus the normalized energy density $\epsilon$ and the amplitude $\mu$ of the
modulation in the off-diagonal hopping. Here we fix the on-site incommensurate potential strength $V=1$, the system size $L=16$ and the number of samples $50$.}
\end{figure}

To indicate that the current critical phase is not due to mobility edge, we show the averaged ratio $\langle r\rangle$ of adjacent gaps versus the normalized energy density $\epsilon$ in Fig. \ref{me}. The rescaled many-body spectrum $\epsilon=\frac{E-E_0}{E_{max}-E_0}$, with $E_0$ ($E_{max}$) being the ground-state (maximum) energy, and the average $\langle r\rangle$ is obtained over the eigenstates in each rescaled energy window $\Delta \epsilon=0.05$ and over samples. From the Fig. \ref{me}, we can clearly see that there is no obvious mobility edges when $\mu>\mu_c$ ($\mu_c\simeq1$ is the ergodic-MBC phase transition point). Accordingly, our study by considering the mid one-third states is reasonable.

\section{II. Unfolding procedure}

\subsubsection{A. Unfolding procedure}
We have studied the level statistics in the main text by unfolding the spectrum. We describe the unfolding procedure here. For an ordered sequence of eigenvalues $\{E_1, E_2, \cdots ,E_M\}$, we can define a stick spectral function:
\begin{equation}
D(E)=\sum_{m=1}^{M}\delta(E-E_m).
\end{equation}
To unfold this spectrum, we further define a cumulative spectral function $\eta$,
\begin{equation}
\eta(E)=\int^{E}_{-\infty}D(E^{'})dE^{'}=\sum_{m=1}^{M}\Theta(E-E_m).
\end{equation}
This function counts the number of eigenvalues less than and equal to E and it is a staircase function, which can be decomposed into a smooth part $\bar{\eta}(E)$ and a fluctuating part $\delta\eta(E)$: $\eta(E)=\bar{\eta}(E)+\delta\eta(E)$.

Unfolding procedure corresponds to mapping the eigenvalues to the
smooth part, $E_n\rightarrow \bar{\eta}(E_n)$, i.e.,
mapping the sequence $\{E_1, E_2, \cdots , E_M\}$ to the numbers $\{\bar{\eta}(E_1), \bar{\eta}(E_2),\cdots, \bar{\eta}(E_M)\}$.
As required, the density of state of the unfolded spectrum, i.e. the derivative of the smooth part, is unity.

In practice, the separation of a spectrum into a smooth part and a fluctuating part is not a trivial task. There are many ways of unfolding \cite{GGS}. After making some tests and comparisons, the method we used is using a polynomial regression of degree
$3$ to fit the staircase function in the range we considered \cite{GGS}, which can keep
the correlations at both short-range and long-range.

\subsubsection{B. Number variance with different sizes}
\begin{figure}[t]
\includegraphics[width=0.6\textwidth]{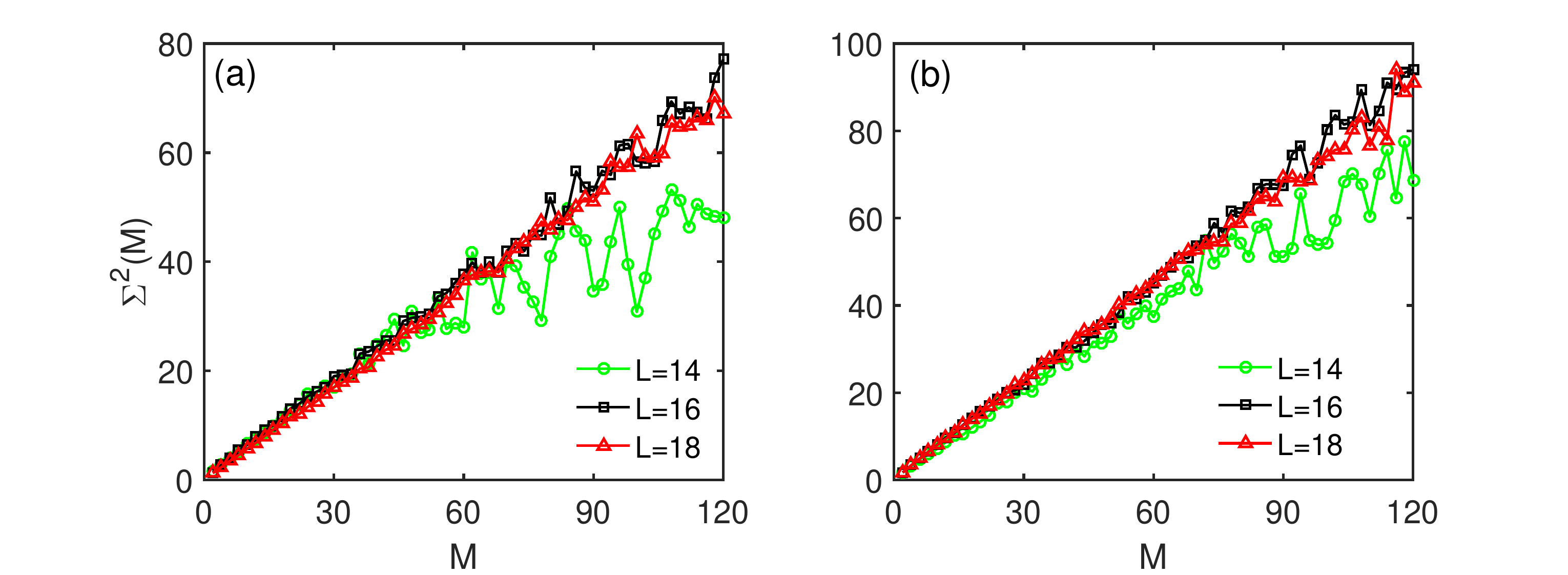}
\caption{\label{S2}
 Number variance $\Sigma^2(M)$ for different system sizes with fixed (a) $\mu=2, V=1$ (in the critical regime) and (b) $\mu=0.5, V=3$ (at the ergodic-MBL transition point).}
\end{figure}

In Fig.2 of the main text, we display the number variance $\Sigma^2(M)$ as a function of $M$ with fixed $L=16$. To see how the slope of the number variance for the MBC phase change with the system size, we show the number variance average over the mid
one-third energy levels of the spectrum and different samples for different sizes in Fig. \ref{S2}. As can be seen, $\Sigma^2(M)$ are virtually size-independent when $M$ is small and it is linear $\Sigma^2(M)\sim \chi M$ with the slope $\frac{1}{2}< \chi< 1$. When $M$ is bigger, the number variance with $L=14$ deviates from the linear relation and shows obvious fluctuations. Further increasing $M$, the number variance with $L=16$ also shows fluctuations. This is because
the Hilbert space size is small for the small system size, which gives a larger statistical error for a bigger $M$.
Thus we have chosen the truncation at $M=100$ in the main text and the slope of the number variance will remain unchanged when increasing the system size.

\section{III. Scaling analysis}

\subsubsection{A. Scaling analysis}
The transition points $V_c$, $\mu_c$ and the corresponding critical exponents $\nu$ can be estimated using a scaling analysis on the average half-chain entanglement entropy $\langle S\rangle$. The angular bracket denotes the average over the mid one-third states and over different samples. $\langle S/S_T\rangle$, where $S_T$ is the Page value $S_T = 0.5[L\ln 2-1]$, can be written as a function of $(P-P_c)L^{1/\nu}$, where $P$ denotes $V$ or $\mu$ and $P_c$ denotes $V_c$ or $\mu_c$.
$P_c$ and the corresponding $\nu$ can be determined by minimizing the quantity \cite{GGS}:
\begin{equation}
Sr(P_c,\nu)=\frac{1}{P_{max}-P_{min}}\int^{P_{max}}_{P_{min}}Var_L\{\langle S/S_T\rangle[(P-P_c)L^{1/\nu}]\}dP,
\end{equation}
where $Var_L$ is the variance over different sizes L. For different sizes, we need use the equal $(P-P_c)L^{1/\nu}$ to calculate the variance for each $P$, so the corresponding $\langle S/S_T\rangle$ need to be interpolated using cubic splines. Where $P_{min}$, $P_{max}$, $P_c$ and $\nu$ can be extracted in an appropriate range to obtain $P_c$, $\nu$ and their errors. For the case of fixing $\mu=0.5$, as shown in Fig.2(a) in the main text, $V_{min}$ is extracted from a box distribution between $2$ and $2.5$, and $V_{max}$ between $3.5$ and $4$. $V_c$ and $\nu$ are taken from a box distribution $[2.5, 3.5]$ and $[0.1, 1]$ respectively. For the selected $V_{min}$, $V_{max}$, $V_c$ and $\nu$, $Sr$ can be obtained. When fixing $V_{min}$ and $V_{max}$, the minimum value of $Sr$ provide the workable $V_c$ and $\nu$. Choosing different $V_{min}$ and $V_{max}$ from the corresponding box distributions, one can obtain different workable $V_c$ and $\nu$, which can provide the errors in $V_c$ and $\nu$. For other parameters, one can obtain the $V_c$, $\mu_c$, the corresponding $\nu$ and their errors by using the same method.

\subsubsection{B. Entanglement entropy with system sizes $L=20, 22$}
In the main text, we show the entanglement entropy (EE) with system sizes from $L=12$ to $L=18$. To further ensure that our results can generalize to bigger sizes, we here also show the data with $L=20$ and $L=22$ by using the shift invert diagonalization
method~\cite{Alet2018}, which is the state-of-the-art method in studying MBL with sizes $L\geq 20$. In this method, we shall consider the mid 100 states for
$L = 20, 22$.

\begin{figure}
\centering
\includegraphics[width=0.6\textwidth]{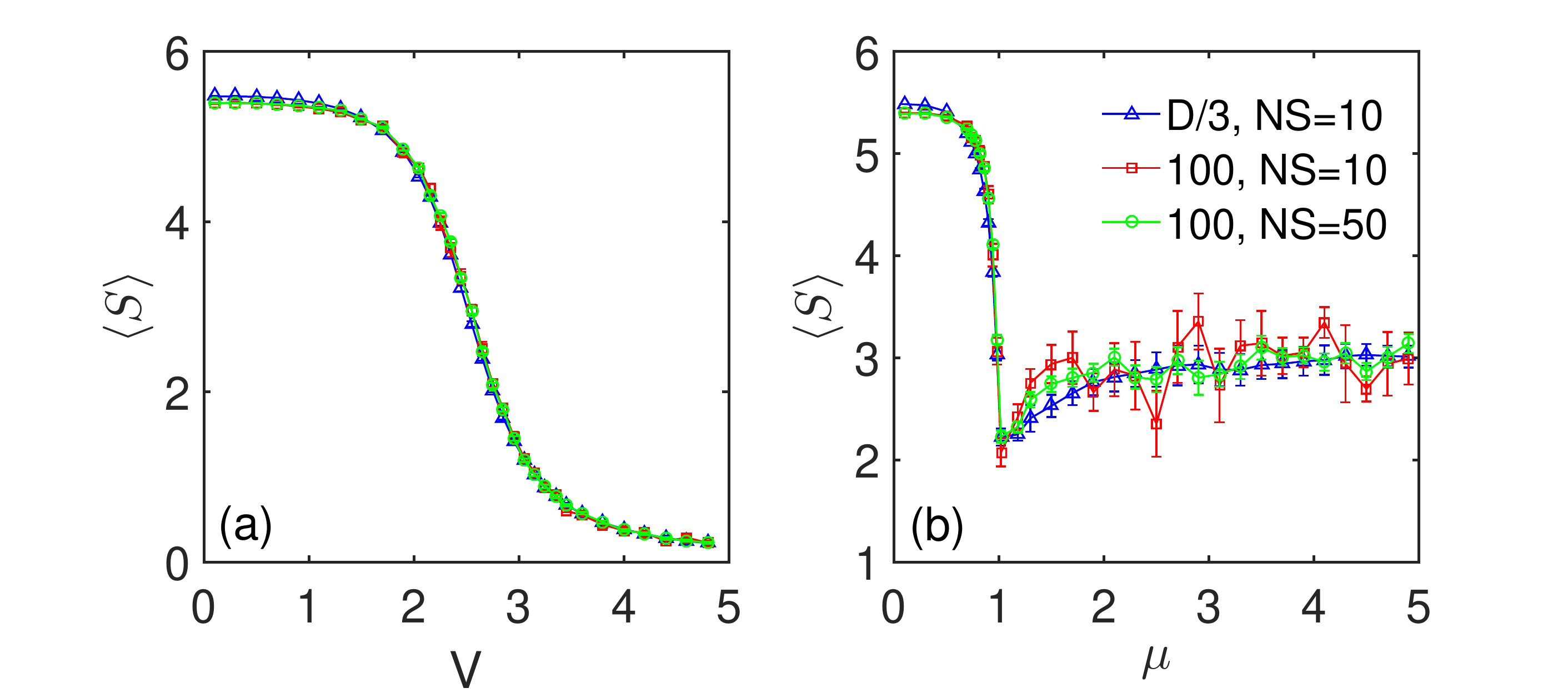}
\caption{\label{EEs}
 The average EE $\langle S\rangle$ versus (a) $V$ with fixed $\mu=0.5$ and (b) $\mu$ with fixed $V=1$. The $\langle S\rangle$ averaged over two energy windows: the mid $\frac{D}{3}$ states with the number of samples (NS) $10$ and mid $100$ states with the NS $10$ and $50$ respectively. Here the system size used is $L=18$.}
\end{figure}

In general, as shown in Fig.~\ref{EEs}, the EE $\langle S\rangle$ averaged over the middle $\frac{D}{3}$ and $100$ states, even with the same number of samples, have no obvious differences in the ergodic and MBL phases. In comparison, in the MBC phase, the averaged EE have larger fluctuation and error when considering the middle $100$ states. This difference is indeed a reflection of the larger fluctuation of the eigenstates' distribution in MBC phase, and it can be quantitatively explained by the fractal dimension $a$ of each eigenstate $|n\rangle$ as $a=-\ln \langle\sum_j|\psi_j|^4\rangle/\ln D$, with different eigenstates giving different fractal dimension in MBC phase. For the ergodic (or MBL) phase, $a$ tends to $1$ (or $0$) for nearly all states, but for the MBC phase, $a$ fluctuates in a reasonable range explicitly different from $0$ and $1$, giving the relatively larger fluctuation of EE.
To reduce the numerical error, we take a relatively large number of samples for this
calculation. We show that the consideration of more samples is reliable from the results presented in
Fig.~\ref{EEs}, where the numerical calculation of EE for $L = 18$ is illustrated by taking the middle one-third
part (i.e. $D/3$) of all the states with small sample number, and the middle $100$ states with a relatively
large number of samples. It can be seen that in the ergodic extended phase and MBL phase, the two
choices well match each other. For the MBC phase, the later choice tends to give the same result of
the former choice by increasing the sample number.


\begin{figure}
\centering
\includegraphics[width=0.6\textwidth]{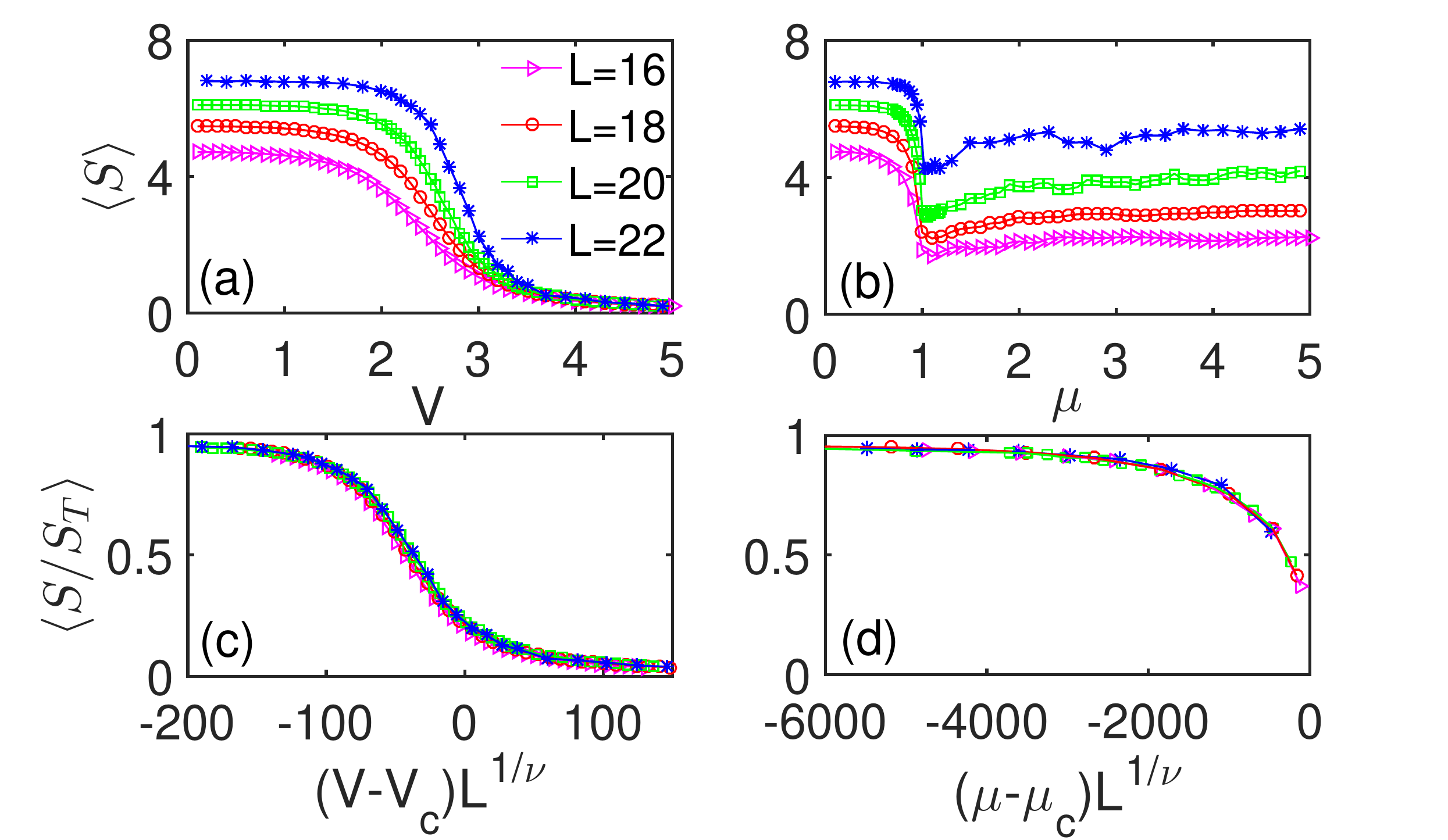}
\caption{\label{EESn}
 (a) The sample averaged EE $\langle S\rangle$ as a function of V with fixed $\mu=0.5$ and (b) $\langle S\rangle$ as a function of $\mu$ with fixed $V=1$. Finite-size scaling analysis of $\langle S/S_T\rangle$ as a function of $(V-V_c)L^{1/\nu}$ with fixed $\mu=0.5$ in (c) and $(\mu-\mu_c)L^{1/\nu}$ with fixed $V=1$ in (d). Rescaled $\langle S/S_T\rangle$ so that all curves for different sizes converge to a single curve.
  Here the number of the samples used are $50$ ($L=16$), $10$ ($L=18$), $100$ ($L=20$) and $50$ ($L=22$). We consider the mid $\frac{D}{3}$ states for $L=16, 18$ and consider the mid $100$ states for $L=20, 22$.}
\end{figure}

Different energy windows have little effect on the EE in the ergodic and MBL phases, so we can consider the average EE $\langle S\rangle$ averaged over the mid one-third states for $L=16, 18$ and over the mid $100$ states for $L=20, 22$, as showed in Fig. \ref{EESn}.
From Fig. \ref{EESn} (a), one can see that the average EE change from a volume law to an area law with the phase transition from the ergodic phase to the MBL phase. The Fig.~\ref{EESn} (b) shows that the average EE follows a volume law when the system is the critical phase.
The EE in the ergodic phase should be the maximum EE, i.e.,$\langle S\rangle\sim \ln\Omega=\ln2^{\frac{L}{2}}=\frac{L}{2}\ln 2$, where $\Omega$ is the total number of states of the half-chain. The average EE in the MBC phase, even though just considering the mid 100 engenstates, are obviously less than the maximum EE, so the mid 100 eigenstates are extended but non-ergodic.
Fig.~\ref{EESn} (c) and (d) show that the EE are fit to $\langle S/S_T\rangle=f[(V-V_c)L^{1/\nu}]$ with fixed $\mu=0.5$, and $\langle S/S_T\rangle=g[(\mu-\mu_c)L^{1/\nu}]$ with fixed $V=1$, respectively. Here $S_T$, $V_c$ and $\mu_c$ denote the Page value, transition point from the ergodic phase to MBL phase and MBC phase, respectively, and $\nu$ is the associated critical exponent.
The results of the best fit are (c) $V_c = 3.08 \pm 0.08$ and $\nu = 0.71 \pm 0.06$ and (d) $\mu_c = 1.03 \pm 0.05$ and $\nu = 0.32 \pm 0.03$, which are consistent with the results in the main text.

In addition, in Fig.4 (d) of the main text, Fig. \ref{S1} (d) and Fig.~\ref{EESn} (d), we only plot out rescaled data in the thermal regime. This is because when $\mu >\mu_c$, this
system enters the critical phase. As we have showed in Eq.(3) of the main text, the critical exponents corresponding to the transition from the ergodic to MBC phases and the transition from the MBC to MBL phases are different. Actually, different points in the critical region give different critical exponents, so the curves for
different sizes in the critical phase can't converge to a single curve with a fixed $\nu$. Thus, if a system has a critical phase, the finite size scaling analysis should be performed over the parameters range from the upper critical parameter to larger parameters and range from the lower critical parameter to smaller parameters respectively \cite{Landau,Isakov2003,Wang2}, which is enough to give the transition points.
In order to perform finite size scaling analysis for the critical phase, one should consider more properties and introduce more critical exponents, which should fulfill some hyperscaling laws in the whole critical region~\cite{Landau,Isakov2003,Wang2}. These hyperscaling laws should be also valid
for the critical point between the ergodic and MBL phases.
But according to the present study on MBL and the transition from the ergodic to MBL phases, it is unclear how to introduce other critical exponents. Therefore, we can't perform finite size scaling analysis and find the hyperscaling laws in the MBC region. But it doesn't affect that performing finite size scaling analysis to determine the phase boundaries and the corresponding critical exponents, and thus we can further determine that this interaction system include three fundamental phases: the ergodic phase, the MBL phase and a new phase.

\subsubsection{C. Finite size effect of fractal dimension}
\begin{figure}
\centering
\includegraphics[width=1\textwidth]{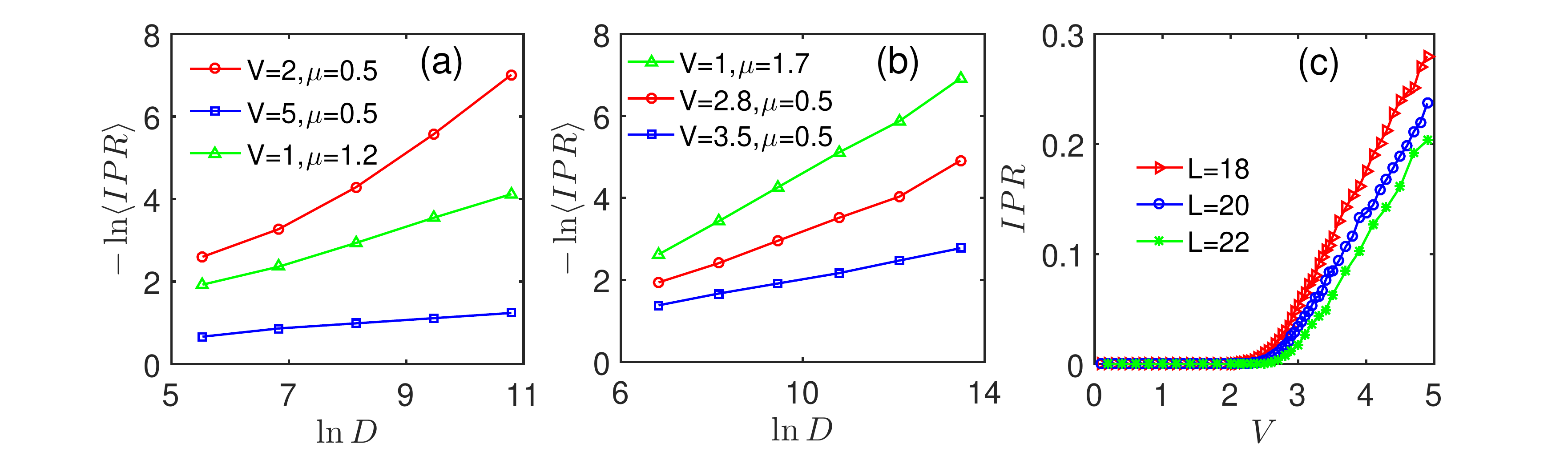}
\caption{\label{mulS}
(a) $-\ln\langle IPR\rangle$ as a function of $\ln D$ with the same parameters of Fig.3(b) in the main text but with smaller system sizes from $L=10$ to $L=18$. We use $-\ln\langle IPR\rangle=a\ln D+b\ln(\ln D)+c$ to fit the data, where $a$, $b$ and $c$ are the coefficients. The
fitting coefficients are $a=1.04\pm0.11$ for $V=2,\mu=0.5$ (in the ergodic phase), $a=0.48\pm 0.07$ for $V=1,\mu=1.2$ (in the MBC phase) and $a=0.09\pm 0.06$ for $V=5,\mu=0.5$ (in the MBL phase). (b) $-\ln\langle IPR\rangle$ as a function of $\ln D$ for different system sizes from $L=12$ to $L=22$. The fitting coefficients are $a=0.64\pm 0.04$ for $\mu=1.7, V=1$ (in the MBC phase), $a=0.53\pm 0.14$ for $\mu=0.5, V=2.8$ (in the ergodic phase, near the ergodic-MBL transition point $V_c$), $a=0.16\pm0.05$ for $\mu=0.5, V=3.5$ (in the MBL phase, near $V_c$). (c) $\langle IPR\rangle$ as a function of $V$ with fixed $\mu=0.5$ for different system sizes. The transition point between the ergodic and MBL phases is $V_c=3.0(8)$ with fixed $\mu=0.5$.
  Here the number of the samples used are $300$ ($L=12$), $200$ ($L=14$), $100$ ($L=16,18,20$), $50$ ($L=22$). We consider the mid $100$ states for all sizes.}
\end{figure}

In this subsection, we show the finite size effect on the fractal dimension $a$ and the inverse participation ratio (IPR), defined in the main text.
In Fig. \ref{mulS} (a), we show $-\ln\langle IPR\rangle$ versus $\ln D$ with the same parameters in Fig.3(b) of the main text but with smaller system sizes.
Contrasting this figure with Fig.3(b), one can see that the fractal dimension $a$ in the MBL phase decreases with increasing the system size.  Thus, it is a consequence of finite-size effect that the fractal dimension $a$ in the MBL is close to zero but not equal to zero.
Further, from the Fig. \ref{mulS} (b), we see that near the transition point, both the MBL phase and the ergodic phase show the multifractal behavior for the system with finite sizes, even though the ergodicity requires $a=1$ in the ergodic phase. From Fig. \ref{mulS} (c), we also see that the averaged IPR changes from zero to non-zero before it reach the transition point from the ergodic to MBL phases.
Therefore, a finite-size system may exhibit a critical region near the transition point between the ergodic and
MBL phases, when the system size is smaller than the correlation length (for ergodic phase) or localization length (for MBL phase).  The critical region vanishes in the thermodynamic limit after performing the finite-size analyses. Increasing the quasiperiodic potential strength $V$ reduces the localization length of wave-functions in MBL phase and thus reduces the finite-size effect. Therefore, as one increases the system size with fixed $V$ or increase $V$ with fixed system size, the fractal dimension $a$ in the MBL phase will decrease. On the other hand, from Fig.\ref{mulS} (a) and (b), we see that $a$ is far from $0$ and $1$ in the whole MBC phase, manifesting that it is deeply multifractal.

On the other hand, all physical results beyond the finite size effect in numerical calculations give the same conclusion: the MBL phase is not multifractal.
In particular, a recent theory developed based on random matrices~\cite{Gao} showed that the MBL states are not multifractal (the fractal dimension $a$ equals zero). Importantly, this theory is not restricted by finite system size, and their result is valid in the thermodynamical limit.
Moreover, as mentioned above, a general energy level spacings distribution is derived as $P(\delta_n)=C_1\delta^{\beta}_nexp(-C_2\delta_n^{2-\gamma})$~\cite{Serbyn2016}, and the Poisson and GOE distribution respectively correspond to $\gamma=1, \beta=0$ and $\gamma=0, \beta=1$. For $IPR\propto D^{-a}$ ($D$ is the Hilbert space size), the fractal dimension $a$ satisfies $a=1-\gamma$~\cite{Serbyn2016}. As we known, level spacings of MBL phase follow the Poisson distribution, where $\gamma=1$, and then $a$ equals zero. We think these analytical results are more convincing compared with numerical calculation in the finite-size systems. Finally,
in the MBL phase the system should remember the local information of an initial state in the dynamical evolution, which also requires the fractal dimension $a$ to be zero. Specifically, we consider an initial state $|\Psi(0)\rangle$, whose density of particles is spatially nonuniform. The state evolves under the MBL Hamiltonian, and we measure the return probability of the initial state after the evolution time $t$, given as $P(t)=|\langle\Psi(0)|\Psi(t)\rangle|^2$, where $|\Psi(t)\rangle$ is the state at time $t$.
By using $|\Psi(t)\rangle=\sum_{n}e^{-iE_nt}|\varphi_n\rangle\langle\varphi_n|\Psi(0)\rangle$, where $|\varphi_n\rangle$ are many-body eigenstates with eigenvalues $E_n$, one can obtain the long-time average of return probability
 $\overline{P}=\lim_{T \rightarrow \infty}\frac{1}{T}\int_0^{T}dt\,P(t)=\sum_{n}|\langle\varphi_n|\Psi(0)\rangle|^4$,
which renders the form proportional to IPR, hence $\overline{P}\sim D^{-a}$. If $a>0$ in the MBL phase, $\overline{P}\rightarrow 0$ in the thermodynamic limit $D\rightarrow\infty$. This contradicts the conclusion that MBL system can preserve the local information of the initial state. This proof shows that the MBL phase should not be multifractal.

\section{IV. Different phases characterized from a one-particle perspective}
\begin{figure}[t]
\includegraphics[width=0.85\textwidth]{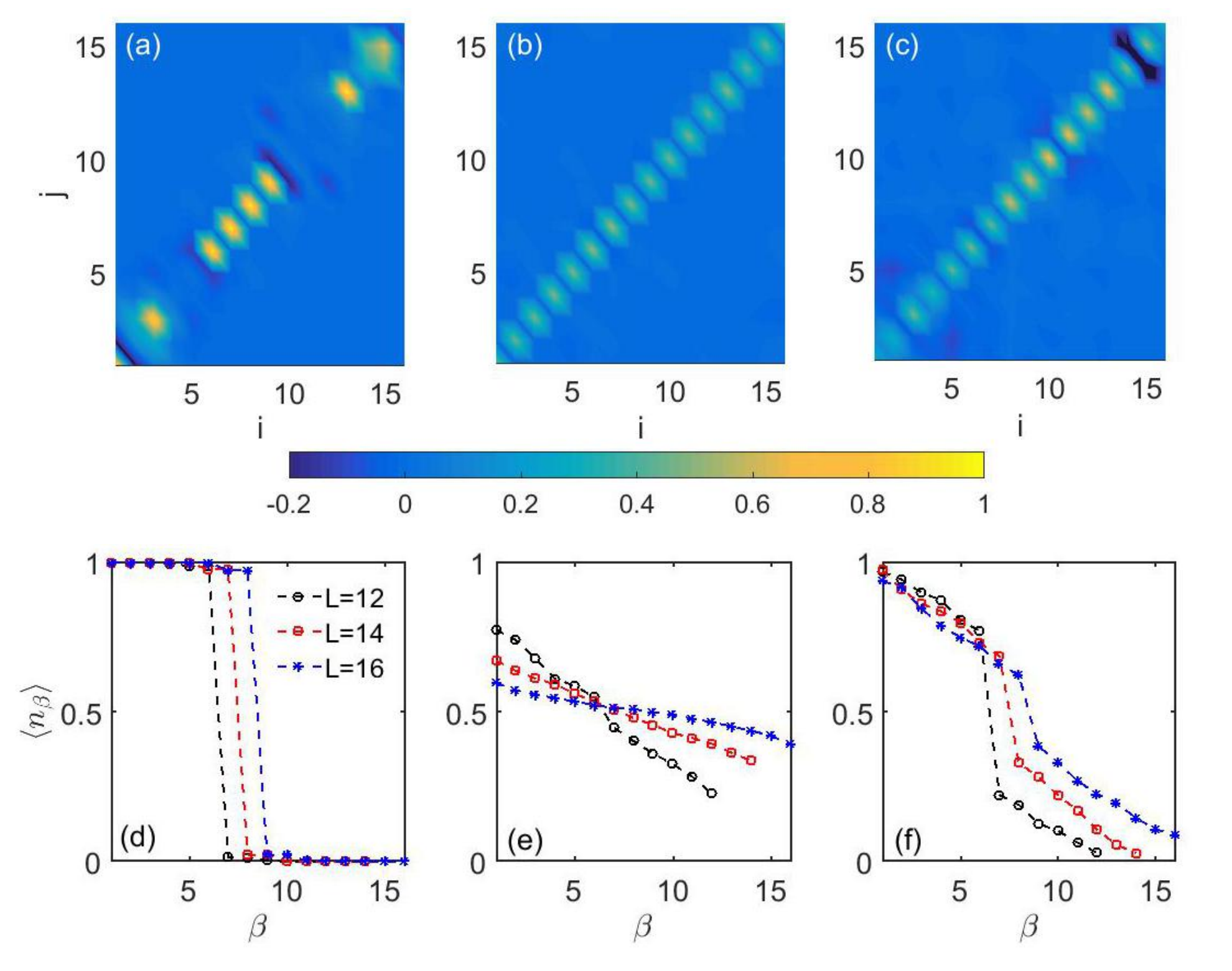}
\caption{\label{S3}
The values of $\rho_{ij}$ versus $i$ and $j$ for (a) $\mu=0.5, V=5$, (b) $\mu=0.5, V=1$ and (c) $\mu=2, V=1$. Here we consider the center state of the spectrum with system size $L=16$ and $\delta=0$.
 (d), (e) and (f) show the averaged occupation of the NOs for different sizes, and their parameters correspond to (a), (b) and (c) respectively.}
\end{figure}

In this section, we use the one-particle density matrix (OPDM) to characterize the different phases~\cite{Bera1,Bera2}. Given a many body wave function $|\psi_m\rangle$, the OPDM is defined as
\begin{equation}
\rho_{ij}=\langle\psi_m|c^{\dagger}_{i}c_j|\psi_m\rangle,
\end{equation}
The diagonalization of the OPDM yields the natural orbitals (NOs) $|\phi_{\beta}\rangle$ with $\beta=1,2,\cdots,L$, and the corresponding eigenvalues $n_{\beta}$ indicating the occupations with $\sum_{\beta=1}^{L}n_{\beta}=N$: $\rho|\phi_{\beta}\rangle=n_{\beta}|\phi_{\beta}\rangle$.
The eigenvalues $n_{\beta}$ are similar to the quasi-local conserved quantities $n^{q}_i$, which can be used to express an effective Hamiltonian for the MBL phase~\cite{Vosk,Serbyn}: $H=\sum_i\epsilon_in^{q}_i+\sum_{ij}J_{ij}n_i^{q}n_j^{q}+\cdots$, where $J_{ij}$ decay exponentially with distance $|i-j|$. When the interaction strength tends to zero, the NOs become the single-particle Anderson orbitals. In the presence of interactions, the NOs can be thought as the dressed versions of the single-particle orbitals, which can be used to characterize the different phases.
Fig. \ref{S3}(a),(b) and (c) directly display each element of the OPDM $\rho_{ij}$, which corresponds to the state of the band center. We observe that in the MBL phase (Fig. \ref{S3}(a)), the diagonal elements are close to zero or close to $1$, implying that half of the NOs are almost fully occupied and the other half are almost unoccupied; in the ergodic phase (Fig. \ref{S3}(b)), all diagonal elements approach $\frac{1}{2}$, signifying that all orbitals are almost occupied equally. In comparison, in the MBC phase (Fig. \ref{S3}(c)), all orbits are occupied, but the probabilities of the occupancies show obvious differences,
implying that the distributions of MBC phase in the NOs show multifractal behaviors.
The same conclusions can be seen more clearly by the eigenvalues $n_{\beta}$, as showed in
Fig. \ref{S3}(d),(e) and (f), where the $n_{\beta}$ are listed in descending order. The angular
bracket denotes the average over different different samples. We see that in the MBL phase, $\langle n_{\beta}\rangle\approx 1$ for $\beta\leq N$, $\langle n_{\beta}\rangle\approx 0$ for $\beta\geq N+1$; in the ergodic phase, increasing the system size makes the occupations more uniform, suggesting that $\langle n_{\beta}\rangle\rightarrow \frac{N}{L}$ for all $\beta$ with $L\rightarrow\infty$. However,
in the MBC phase, the distributions of $n_{\beta}$ is wide, i.e., almost all orbits are occupied with different probabilities, indicating that the MBC state is extended but non-ergodic.

\begin{figure}[t]
\includegraphics[width=0.65\textwidth]{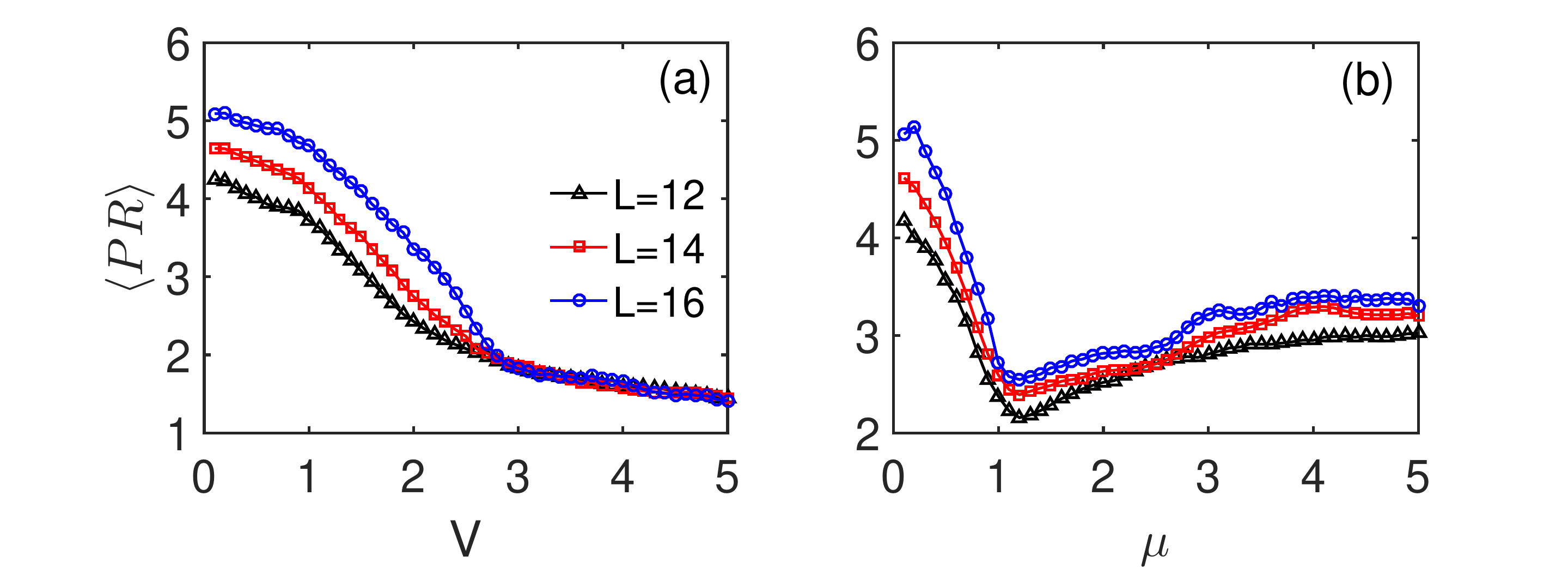}
\caption{\label{S4}
 (a) $\langle PR\rangle$ as a function of $V$ for different sizes with fixed $\mu=0.5$, (b) $\langle PR\rangle$ as a function of $\mu$ for different sizes with fixed $V=1$. The angular bracket denotes the average over different samples and different states, and here consider $10$ eigenstates closest to the band center.}
\end{figure}

Since the NOs can be analyzed in the
similar way as single-particle orbitals, we consider the inverse participation ratio (IPR) from the NOs:
\begin{equation}
IPR=\frac{1}{N}\sum_{\beta=1}^{L}n_{\beta}\sum_{j=1}^L|\phi_{\beta}(j)|^4.
\end{equation}
The average participation ratio (PR) is defined as $\langle PR\rangle=\langle 1/IPR\rangle$, satisfying $\langle PR\rangle\sim L^{\eta}$, with $\eta$ being a fractal dimension.
It has been known that $\eta\rightarrow 0$ ($\eta\rightarrow 1$) for localized (extended) states, while $0<\eta<1$ corresponding to critical states.
Fig. \ref{S4} display the behaviors of $\langle PR\rangle$. In the MBL phase, $\langle PR\rangle$ remains a small size-independent value, suggesting that all occupied states are localized. In both the ergodic and MBC phases, $\langle PR\rangle$ increases with the system size, meaning that they are delocalized. However, in the MBC phase the increase with size is much slower, showing the multifractal feature ($0<\eta<1$) in the MBC phase.

With the above results, we can provide now intuitive picture for the understanding of the MBC phase. As for strong disorder strength, the Anderson localization becomes MBL when adding interactions. In orbital language, it corresponds to that the localized occupations in single particle Anderson orbitals become the localized occupations in the NOs. Similarly, for the critical phase,
the multifractal behavior in single particle orbitals becomes the multifractal distribution of qusi-particles in NOs when the interaction is added.
For this the single-particle critical phase turns into the MBC phase. In particular,
when $\mu\gg U$, the interactions can be treated as perturbation and the NOs can be approximated by the single particle orbitals, so in this case, the MBC phase is naturally stable with the weak interaction. When $\mu$ and $U$ have the same order of magnitude, the above numerical results tell us that the occupations in NOs still show multifractal behavior, hence still in the MBC regime. This is also consistent with the fact that the MBC phase is delocalized and non-ergodic.


\end{document}